\documentclass[aps,pra,showpacs,twoside,twocolumn,nofootinbib,10pt]{revtex4-1}
\usepackage[colorlinks=true, citecolor=red, urlcolor=blue ]{hyperref}
\usepackage{epsfig,newlfont,amssymb,amsfonts,amsmath,bm,subfigure,palatino,mathtools,amsthm,braket,times,soul}
\usepackage{color}
\usepackage{comment}
\usepackage{multirow}
\usepackage{ulem}
\usepackage{physics}
\definecolor{indiagreen}{rgb}{0.07, 0.53, 0.03}
	\definecolor{teal}{rgb}{0.0, 0.53, 0.53}

\begin{document}

\title{
Dimensional advantage in secure information trading via the noisy dense coding protocol
	}

	\author{Ayan Patra\(^1\), Rivu Gupta\(^1\), Tamoghna Das\(^2\), and Aditi Sen(De)\(^{1}\)}
	
	\affiliation{\(^1\)Harish-Chandra Research Institute, HBNI, Chhatnag Road, Jhunsi, Allahabad 211 019, India}
	\affiliation{\(^2\)Department of Physics, Indian Institute of Technology Kharagpur, Kharagpur 721302, India}

 \begin{abstract}


 The quantum dense coding (DC) protocol, which has no security feature, deals with the transmission of classical information encoded in a quantum state by using shared entanglement between a single sender and a single receiver. Its appropriate variant has been established as a quantum key distribution (QKD) scheme for shared two-qubit maximally entangled states, with the security proof utilizing the uncertainty relation of complementary observables and the Shor-Preskill entanglement purification scheme. We present the DC-based QKD protocol for higher dimensional systems and report the lower bounds on secret key rate, when the shared state is a two-qudit maximally entangled state, and mixtures of maximally entangled states with different ranks. The analysis also includes the impact of noisy channels on the secure key rates, before and after encoding. In both the noiseless and the noisy scenarios, we demonstrate that the key rate as well as the robustness of the protocol against noise increases with the dimension. Further, we prove that the set of useless states in the DC-based QKD protocol is convex and compact.  

\end{abstract}
	
	\maketitle

\section{Introduction}
\label{sec:intro}

Quantum cryptography~\cite{Gisin_RMP_2002, Scarani_RMP}, which is based on the combination of quantum mechanics and classical information theory, is advantageous due to the crucial resource of bipartite entanglement~\cite{HoroRMP} or non-classical correlations~\cite{CHSH, Bell-nonlocality}. Any secure quantum key distribution (QKD) protocol has two major paradigms --  device-dependent (DD)~\cite{BB84,Gisin_RMP_2002,AcinBBBMM2004} and device-independent (DI)~\cite{Ekert1991,Mayers-Yao,acin-2007-98, Masanes2011} regimes. The former assumes that the adversary, Eve, is permitted to perform only a quantum mechanically allowed set of operations, and the honest parties, Alice and Bob, require a complete specification of the devices used and the measurements they perform. On the other hand, in the DI scenario, it is sufficient to ensure that the correlation shared by the participants violates local realism~\cite{EPR}, and additional specifications about the internal workings of the devices are not required~\cite{acin-2007-98}. 
Recently, a lot of interest has emerged in the direction of secure key agreement protocols where the laws of quantum mechanics do not restrict Eve and the honest parties, and they can perform any operation bounded only by the laws of no-signaling theory~\cite{Kent,AcinGM-bellqkd, masanes-2006, acin-2006-8, hanggi_2009}.



Bipartite entanglement serves as a quintessential resource in several other novel protocols like superdense coding~\cite{BennettWisener}, teleportation~\cite{Bennett_PRL_1993}, catalytic state transformation~\cite{Jonathan_PRL_1999}, and remote state preparation~\cite{Bennett_PRL_2002} to name a few. In this work, we specifically concentrate on the superdense coding (DC) protocol, which uses a shared entangled state to double the capacity of classical information transmission between a sender and a receiver. Moreover, the quantum advantage is shown to be robust even if noise affects the shared pure state during the state distribution~\cite{Bose, Bowen_PRA_2001, Horodecki_arxiv_2001, Ziman_PRA_2003, Bruss_PRL_2004} or during the communication of the encoded component~\cite{Quek_PRA_2010, Shadman_NJP_2010, Shadman2011, Shadman_PRA_2012, Shadman2013, Das_PRA_2014, Das_PRA_2015, Mirmasoudi_JPA_2018} (for multipartite DC schemes, see Refs. ~\cite{Badziag_PRL_2003, Bruss_PRL_2004, Aditi} ). 
 Interestingly, it has been demonstrated that the shared two-qubit entangled state utilized for the DC protocol can be made secure against a quantum adversary when modified using suitable procedures~\cite{Degivanni_PRA_2004, Beaudry2013, Han_SR_2014}.
 In particular, the sender, Alice, prepares a bipartite quantum state, keeps one part with her in a quantum memory, and sends the other part to the receiver, Bob, with the help of a quantum channel which he again sends back to Alice after encoding (two-way protocol) through the same or some other quantum channel. The main benefit of this QKD scheme is that no basis choice is required, thereby making it deterministic, although the eavesdropper, Eve, can attack the signal twice in this case, once when Alice distributes the state and again when Bob delivers the encoded state back. 
 The security analysis in this situation relies on the fact that any orthogonal unitary operation on an unknown quantum state can be purified to a joint von Neumann measurement along with an auxiliary quantum state ~\cite{Beaudry2013}, and that entropic uncertainty relations~\cite{Coles_RMP_2017} governing the complementary observables are similar to the generalized Pauli $X$ and $Z$ measurements that are essential to the BB84 protocol.
 

On a different front, arbitrary higher dimensional quantum states, qudits, have been demonstrated to outperform their lower dimensional counterparts in a variety of quantum technologies, including quantum switches~\cite{Wei_PRL_2019}, quantum simulation~\cite{Neeley_Science_2009, Kaltenbaek_NP_2010}, robust entanglement distribution~\cite{Ecker_PRX_2019}, quantum computation~\cite{Lanyon_NP_2009, Babazadeh_PRL_2017, Muralidharan_NJP_2017}, quantum telecloning~\cite{Nagali_PRL_2010, Bouchard_SA_2017}, quantum batteries~\cite{Santos_PRE_2019, Dou_EPL_2020, Ghosh_PRA_2022}, and quantum refrigerators~\cite{Correa_PRE_2014, Wang_PRE_2015, Usui_PRA_2021, Konar_PRA_2023}. 
In a similar fashion, it has been shown that QKD using qudits can provide an increased key rate~\cite{Bechmann_PRA_2000, Nikolopoulos_PRA_2006, Sasaki_Nature_2014}, and a higher resilience against noise~\cite{Cerf_PRL_2002, Nikolopoulos_PRA_2005, Doda_PRAppl_2021}, thereby establishing dimensional advantage. 
Moreover, such schemes can be realized in laboratories~\cite{Groblacher_NJP_2006, Zhong_NJP_2015} using twisted photons~\cite{Bouchard_Quantum_2018}, time-bin encoding~\cite{Islam_SA_2017}, telecommunication fibers~\cite{Canas_PRA_2017}, silicon integrated circuits~\cite{Ding_NPJ_2017}, multimode fibers~\cite{Sit_OL_2018, Amitonova_OE_2020} and even underwater~\cite{Xu_OeL_2018, Bouchard_OE_2018}.

In this paper, we present the dense-coding-based two-way QKD protocol by using qudits and exhibit that sharing secure quantum keys in the higher dimensional Hilbert space is beneficial over low-dimensional systems. Specifically, the Bell measurements needed for this protocol can be refurbished in higher dimensions using the Weyl Heisenberg spinors~\cite{Bertlmann_JPA_2008}, and this allows one to present the lower bound on the key rate in arbitrary dimensions. We demonstrate the gain using shared probabilistic mixtures of maximally entangled states of different ranks, and we develop a connection between the information transmission rate in the DC protocol and the secret key rate. 
When both of the quantum channels involved in this QKD routine are noisy, we also estimate the lower bound of the key rate and find the critical noise level up to which a nonvanishing key rate can be achieved. For illustration, three paradigmatic channels - the dit-phase-flip, the depolarising, and the amplitude-damping channels are considered.  We contend that increasing the dimension not only improves the secret key rate but also contributes to its robustness against noise. We prove the closed and compact features of the set of states that are unsuitable for the QKD protocol under scrutiny.


The paper is organized in the following manner. In Sec.~\ref{sec:noiseless_high}, we present the dense-coding-based QKD protocol and the methodology to obtain the key rate.  The connection between key rates and the DC capacity is obtained for pure and mixtures of maximally entangled states in Sec.~\ref{sec:pure-SDC}. After the action of noisy channels, the protocol and its corresponding key rates are presented in Sec.~\ref{sec:noisy_qkd}. Sec.~\ref{sec:set}  proves the properties of the set of states which are not useful for this QKD protocol and the concluding remarks are added in Sec.~\ref{sec:conclu}. 

\section{Secure dense coding-based key distribution protocol in higher dimension }
\label{sec:noiseless_high}

The key distribution scheme~\cite{Bostrom_PRL_2002} based on super dense coding, also known as secure dense coding (SDC), is a two-way quantum cryptography protocol proposed for an entangled two-qubit state shared between two honest parties,  Alice and Bob. The assumptions used to prove its security are semi-device-independent in nature. In this section, we present the dense coding-based key distribution scheme involving higher dimensional systems. Before going into the qudit SDC scheme, let us describe the protocol for an arbitrary two-qubit entangled state, $\rho_{AB}$,  shared between Alice and Bob. The protocol proceeds according to the following steps:  

\begin{enumerate}
    \item Initially, Alice prepares a two-qubit state, represented by $\rho_{AA'}$. She sends qubit $A'$ to Bob via a quantum channel $\mathcal{E}_{A' \to B}$, thereby sharing $\rho_{AB}$, i.e., $\mathcal{E}_{A' \to B}(\rho_{AA'}) = \rho_{AB}$.
    
    \item Bob then operates local unitaries, $\{U^{xy}\}$, on his qubit to encode the message which occurred with probability $p^{xy}$. \textcolor{black}{Here, $x,y \in \{0,1\}$ (i.e., there are four possible unitary operators) and the set of unitaries, $\{U^{xy}\}$, denote the qubit identity and Pauli matrices, i.e.,   $\mathbb{I} (x = y = 0), \sigma_z (x = 1, y = 0), i\sigma_y (x = 1, y = 1)$, and $\sigma_x (x = 0, y = 1)$}, when the shared state is the maximally entangled Bell state, $\rho^{\phi^+}_{AB}$, with $\ket{\phi^+} = \frac{1}{\sqrt{2}} (\ket{00} + \ket{11})$. This leads to the creation of the ensemble, $\{p^{xy}, \rho_{AB}^{xy}\}$, with $\rho_{AB}^{xy} = (\mathbb{I}_A \otimes U^{xy}_{B}) \rho_{AB} (\mathbb{I}_A \otimes U^{\dagger xy}_{B})$.
    
    \item Upon encoding, the qubit is sent back to Alice through another channel $\mathcal{E}_{B \to A'}$, thereby resulting in the ensemble $\{p^{xy}, \rho_{AA'}^{xy} \}$. This step can be represented as $\mathcal{E}_{B \to A'}(\rho_{AB}^{xy}) = \rho_{AA'}^{xy}$.
    
    \item Alice then performs a Bell measurement on both the qubits to recover the message sent by Bob. Note that Bell measurement is the optimal measurement when the shared state is maximally entangled. In case of any arbitrary shared state, we still fix the measurements to be the same.
    
\end{enumerate} 

 As mentioned before, Eve, in this scheme, has two opportunities to attack - once before encoding and again after the encoding process. To ensure security, Alice and Bob perform certain test runs, with a very low probability, to determine the presence of Eve. The raw key generation run, as described above, takes place with almost unit probability. The test runs comprise the steps enumerated below: 
 \begin{enumerate}
     \item Bob measures his qubit of $\rho_{AB}$ in the $\sigma_z$ basis. If the outcome is $\ket{0 (1)}$, he sends back the state $\ket{+(-)}$ where $\ket{\pm} = \frac{1}{\sqrt{2}}(\ket{0} \pm \ket{1})$ through the channel $\mathcal{E}_{B \to A'}$. 

     \item Upon receiving Bob's input, Alice measures her stored qubit in the $\sigma_z$ basis, and the received one in the $\sigma_x$ basis. If her two-qubit measurement outcome corresponds to either $\ket{0+}_{AA'}$ or $\ket{1-}_{AA'}$, she confirms the protocol to be secure. 
 \end{enumerate}

 \textcolor{black}{It is important to note that, since the protocol involves multiple uses of a channel, the question of maintaining its stability for a longer duration of time naturally arises. In a two-way protocol, there can be two distinct scenarios involving the use of channels to transmit the qudit back-and-forth between the parties. In the first case,  independent channels, i.e., identical but separate channels, are employed before and after the encoding process, while in the second situation, the same channel is used twice, which corresponds to the notion of a correlated channel. This work deals with the channels that are identical but separate (independent), and hence, each channel is used for the same duration as in a one-way protocol. Thus, the question of a longer stabilization of the channel becomes irrelevant in this case. On the other hand, for correlated channels, it has been established that the decoherence effect can be mitigated by the use of the same channel, thereby leading to a higher key rate~\cite{Beaudry2013}. However, since the same channel is employed for a longer duration, its stability is a matter of concern.}
 
 In the following discussion, we elucidate the SDC protocol when the shared state comprises two qudits.
 
 \subsubsection{The key generation run}
 \label{subsubsec:key_run-SDC}


Let us present the protocol when the shared state involves qudits (quantum states of dimension $d \geq 3$). Note first that the unitary encoding in higher dimensions is performed through Weyl operators, given by~\cite{Hiroshima_JPA_2001}
\begin{equation}
    U^{xy} = \sum_{l = 0}^{d-1} e^{\frac{2 \pi i}{d} l y} |l \rangle \langle l\oplus x|
    \label{eq:unitary}
\end{equation}
where the symbol $``\oplus''$ denotes sum modulo $d$ and each $(x,y)$-pair corresponds to a particular message that Bob wishes to encode. This can be equivalently represented by a measurement scheme involving an auxiliary maximally entangled state at Bob's side~\cite{Beaudry2013}. In the $d^2$-dimensional complex Hilbert space, $\mathbb{C}^d \otimes \mathbb{C}^d$ (or equivalently $d \otimes d$), the maximally entangled state and the Bell bases for measurement respectively read as
\begin{eqnarray}
   && |\phi^+\rangle_{B' C} = \frac{1}{\sqrt{d}} \sum_{p = 0}^{d-1} |p,p\rangle_{B' C}, \label{eq:high_d_max_ent}\\
   \text{and} ~ &&  |\mathcal{B} (x y)\rangle_{BB'} = \frac{1}{\sqrt{d}} \sum_{l = 0}^{d-1} e^{\frac{2 \pi i}{d} l y} | l, l \oplus x \rangle_{BB'}.\label{eq:high_d_Bell}
\end{eqnarray}
More precisely, the state $\rho_{AB}$ is concatenated with a maximally entangled pure state $|\phi^+\rangle_{B' C}$ whereafter, a Bell-basis measurement, $\{|\mathcal{B}(xy)\rangle\}$, is performed on the parties $BB'$. 
Here, $(xy)$ denotes the measurement outcome with $x, y  = 0,1,2, \dots , d-1$. Each of the $d^2$ measurement outcomes yields a certain encoded message, denoted as $\rho_{A C}^{xy}$. Mathematically, this step can be represented as
\begin{widetext}
\begin{eqnarray}
   _{BB'}\langle \mathcal{B} (xy) | \rho_{AB} \otimes |\phi^+ \rangle_{B' C} \langle \phi^+| \mathcal{B}(xy)\rangle_{BB'} \sim (\mathbb{I}_A \otimes U^{xy}_C ) \rho_{A C} (\mathbb{I}_A \otimes U^{\dagger xy}_C ) \equiv \rho_{A C}^{xy}.
   \label{eq:measure_encode}
\end{eqnarray}
\end{widetext}
Eventually, Bob sends the encoded qudit back to Alice through a channel $\mathcal{E}_{C \to A'}$. Alice's job is to then perform an optimal measurement on $\rho_{AA'} = \mathcal{E}_{C \to A'}(\{ p^{xy}, \rho_{A C}^{xy}\})$ such that both the accessible information and the key rate are maximized. Note that for a shared maximally entangled state, the optimal measurement corresponds to that in the high-dimensional Bell basis. We will first establish Eq.~\eqref{eq:measure_encode}.

Let us consider a two-qudit state, $\rho_{AA'}$, such that 
\textcolor{black}{
\begin{equation}
    \mathcal{E}_{A' \to B}(\rho_{AA'}) = \rho_{A B} = \sum_{j,k,m,n = 0}^{d-1} \alpha_{jkmn} | j,k \rangle_{A B} \langle m,n |,
    \label{eq:rho_AX}
\end{equation}
}
where $|j\rangle, |k\rangle, |m\rangle$, and $|n \rangle$ represent computational basis vectors in the $d$-dimensional Hilbert space. Furthermore, $\alpha_{jkmn}$ are the complex coefficients characterizing the state. Thus, the process of making a Bell measurement on the composite state of $\rho_{A B}$ and $|\phi^+\rangle_{B' C}$ can be explicitly written as 
\begin{widetext}
    \begin{eqnarray}
        && \nonumber  _{BB'}\langle \mathcal{B} (xy) | \rho_{AB} \otimes |\phi^+ \rangle_{B' C} \langle \phi^+| \mathcal{B}(xy)\rangle_{BB'} = \frac{1}{d^2} \sum_{\substack{j,k,m,n,\\p,q,l,l' = 0}}^{d-1} e^{\frac{2 \pi i}{d} y (l' - l)} \alpha{_{jkmn}} _{BB'}\langle l, l \oplus x | j,k,p,p \rangle_{ABB'C}\langle m,n,q,q | l', l' \oplus x \rangle_{BB'} \\
        &&  = \frac{1}{d^2} \sum_{\substack{j,k,m,n,\\p,q,l,l' = 0}}^{d-1} e^{\frac{2 \pi i}{d} y (l' - l)} \alpha_{jkmn} |j,p \rangle_{A C} \langle m,q| \delta_{k,l} \delta_{p, l \oplus x} \delta_{n,l'} \delta_{q, l' \oplus x} = \frac{1}{d^2} \sum_{\substack{j,m,\\l,l' = 0}}^{d-1} e^{\frac{2 \pi i}{d} y (l' - l)} \alpha_{jlml'} |j, l \oplus x \rangle_{A C} \langle m, l' \oplus x |. \label{eq:Bell_m} 
    \end{eqnarray}
\end{widetext}
 The action of the unitaries given by Eq.~\eqref{eq:unitary} on the state $\rho_{AC}$ can be shown to be
\begin{widetext}
\begin{eqnarray}
   && \nonumber (\mathbb{I}_A \otimes U_{C}^{xy}) \rho_{A C} (\mathbb{I}_A \otimes U_{C}^{\dagger xy}) = \sum_{\substack{j,k,m\\n,l,l' = 0}}^{d-1} e^{\frac{2 \pi i}{d} y (l' - l)} \alpha_{jkmn} |l \oplus x \rangle_{C} \langle l|j,k \rangle_{A C} \langle m,n | l' \rangle_{C} \langle l' \oplus x| \\
   && =  \sum_{\substack{j,k,m\\n,l,l' = 0}}^{d-1} e^{\frac{2 \pi i}{d} y (l' - l)} \alpha_{jkmn} |j, l \oplus x \rangle_{A C} \langle m, l' \oplus x | \delta_{k,l} \delta_{n,l'} = \sum_{\substack{j,m\\l,l' = 0}}^{d-1} e^{\frac{2 \pi i}{d} y (l' - l)} \alpha_{jlml'} |j, l \oplus x \rangle_{A C} \langle m, l' \oplus x |. \label{eq:unitary_op}
\end{eqnarray}
\end{widetext}
Thus, comparing Eqs.~\eqref{eq:Bell_m} and~\eqref{eq:unitary_op}, we observe that the concatenation of the shared state with a maximally entangled pure state, followed by a Bell measurement is equivalent to a unitary operation on the original resource with a probability $1/d^2$. Upon completion of the protocol, the resulting classical-classical (cc) state may be represented as
\begin{eqnarray}
    \kappa_{AA'BB'}  = \sum_{i,j,x,y} q(ijxy) |i j\rangle_{AA'} \langle i j| \otimes |xy\rangle_{BB'}\langle x y|,
    \label{eq:cc-state}
\end{eqnarray}
where $|i j\rangle_{AA'} \langle i j|$ represents the register containing Alice's final Bell measurement outcomes, and similarly $|x y\rangle_{BB'} \langle x y|$ for Bob. The corresponding joint probability distribution is given by $q(ijxy)$.  

\subsubsection{The key rate}
\label{subsubsec:key rate-SDC}

The presence of an eavesdropper, Eve (E), in general, makes the shared state a mixed one. One can consider that it is Eve who initially prepares a pure state and distributes two parties to Alice and Bob. By considering a purification of the state as $\ket{\psi}_{AA'BB'E}$, fewer assumptions are required to derive the secure key rate~\cite{Beaudry2013}. \textcolor{black}{After} the key-generation run, the  classical-classical-quantum (ccq) state results in 
\begin{eqnarray}
  && \nonumber \kappa_{AA'BB'E} = \sum_{i,j,x,y} \hat{\mathcal{B}}_{AA'}(ij) \otimes \hat{\mathcal{B}}_{BB'}(xy) (|\psi\rangle_{AA'BB'E} \langle \psi|) \\
    && = \sum_{i,j,x,y} q(ijxy) |i j\rangle_{AA'} \langle i j| \otimes |xy\rangle_{BB'}\langle x y| \otimes \rho_{E}^{ijxy},
    \label{eq:ccq-state}
\end{eqnarray}
where Alice and Bob perform Bell measurements. And the \textcolor{black}{terms, $\hat{\mathcal{B}}(.,.)$, are the superoperators for the Bell measurements, given by $ \hat{\mathcal{B}}(xy) (\rho) = \ket{\mathcal{B}(xy)}\bra{\mathcal{B}(xy)} \rho \ket{\mathcal{B}(xy)}\bra{\mathcal{B}(xy)} $ for all $x,y \in \{0,1\}$}, 
and $\rho_{E}^{ijxy}$ [with probability $q(ijxy)$] is the state in possession of Eve after the protocol is completed. Since $\rho_{E}^{ijxy}$ depends on the measurement statistics of both Alice and Bob, Eve can gather information about the generated key through measurement on the state. The secret key rate, $r$, even in the presence of Eve, is lower bounded by~\cite{Devetak_PRS_2005}
\begin{equation}
    r \geq I(A:B)_{\kappa} - I(B:E)_{\kappa},
    \label{eq:key rate}
\end{equation}
where the mutual information $ I(A:B) \equiv I(\sigma_{AB})  =  S(\sigma_A) + S(\sigma_B) - S(\sigma_{AB})$ with \(S(*) = -\Tr (* \log_2 *)\) being the von Neumann entropy computed with respect to the ccq state in Eq.~\eqref{eq:ccq-state}. Let us recall that  $I(A:B)_\rho = S(B)_\rho - S(B|A)_\rho$ for any bipartite state $\rho_{AB}$, with $S(\rho_A) \equiv S(A)_\rho = -\Tr (\rho_A \log_2 \rho_A)$ as the von Neumann entropy of the reduced system, $\rho_A = \Tr_B(\rho_{AB})$, and $S(B|A)_\rho = S(AB)_\rho - S(A)_\rho$ as the conditional entropy~\cite{nielsenchuang_CUP_2000}. Note that throughout the manuscript, the arguments of the entropy function are specified as the state labels, instead of the states themselves, with the state as the subscript. To derive the secret key rate, let us define two states,
\begin{eqnarray}
   && \nonumber \sigma_{AA'BB'E} = \sum_{i,j,x,y} \hat{\mathcal{T}}_{AA'}(ij) \otimes \hat{\mathcal{B}}_{BB'}(xy) (|\psi\rangle_{AA'BB'E} \langle \psi|), \\
   \label{eq:sigma_state}\\
  && \nonumber  \tau_{AA'BB'E} = \sum_{i,j,x,y} \hat{\mathcal{T}}_{AA'}(ij) \otimes \hat{\mathcal{T}}_{BB'}(xy) (|\psi\rangle_{AA'BB'E} \langle \psi|), \\ \label{eq:tau_state}
\end{eqnarray}
where $\hat{\mathcal{B}}(xy)$ represents the Bell basis measurement discussed above and in Eq.~\eqref{eq:high_d_Bell}. \textcolor{black}{$\hat{\mathcal{T}}(ij)$ is a test measurement, given by $\hat{\mathcal{T}}_{AA'}(ij)(\rho) = \ket{i,j_\vdash}\bra{i,j_\vdash}\rho \ket{i,j_\vdash}\bra{i,j_\vdash}$ for all $i,j \in \{0,1\}$, and is performed in order to detect the presence of Eve, which will be specified later (see Eq.~\eqref{eq:MUB})}. We can readily see that $S(B|E)_\kappa = S(B|E)_\sigma$ since the only difference between the two states in Eqs.~\eqref{eq:ccq-state} and~\eqref{eq:sigma_state} is the measurement performed by Alice, which does not affect the conditional entropy term.

The secret key rate of a key distribution protocol has been derived using entropic uncertainty relations in several previous works~\cite{Cerf_PRL_2002, Grosshans_PRL_2004, Koashi_JP_2006}. We shall now recourse to the entropic uncertainty relation formulated in Ref.~\cite{Berta_NP_2010}: \textcolor{black}{Corresponding to two different POVM settings, denoted as $P_X \equiv \{P^i_X\}_i$ and $P_Z \equiv \{P^j_Z\}_j$,} with classical outcomes, $i$ and $j$ respectively, performed by the party A on a tripartite state $\rho_{ABE}$, the entropic uncertainty relation states

\begin{equation}
    S(Z|B) + S(X|E) \geq \log_2 \frac{1}{c},
    \label{eq:ent_unc}
\end{equation}
where $c = \max_{i,j} || \sqrt{P^i_X} \sqrt{P^j_Z}||_{\infty}^2$, with $i$ and $j$ denoting the measurement outcomes of $P_X$ and $P_Z$, respectively, and $||.||_{\infty}$ stands for the infinity norm~\cite{Watrous_CUP_2018}. Applying the relation to Eqs.~\eqref{eq:sigma_state} and~\eqref{eq:tau_state}, we obtain
\begin{equation}
    S(B|E)_\sigma + S(B|A)_\tau \geq \log_2 \frac{1}{c}.
    \label{eq:entropic_unc_key}
\end{equation}
Here $c = \max_{(ij),(xy)} || \sqrt{\mathcal{T}(ij)} \sqrt{\mathcal{B}(xy)}||_{\infty}^2$. The lower bound on $r$, along with Eq.~\eqref{eq:entropic_unc_key}, becomes
\begin{eqnarray}
   \nonumber && r \geq I(A:B)_\kappa - I(B:E)_\kappa \\
    \nonumber && = S(B|E)_\kappa - S(B|A)_\kappa = S(B|E)_\sigma - S(B|A)_\kappa\\
    && \geq \log_2 (1/c) - S(B|A)_\tau - S(B|A)_\kappa. \label{eq:key_rate_bound}
\end{eqnarray}
If the measurements performed by Alice and Bob to obtain $\kappa_{AA'BB'E}$ and $\tau_{AA'BB'E}$ are fully correlated, which is the case when the shared state is maximally entangled, the last two von Neumann entropy terms in Eq.~\eqref{eq:key_rate_bound} vanish and thus $r \geq \log_2(1/c)$.
It can be shown that the secret key rate is optimal when the test measurement conducted by Alice to obtain $\tau_{AA'BB'E}$ comprises the following mutually unbiased set of bases~\cite{Das_arXiv_2021}
\begin{equation}
\mathcal{T} \equiv  |j,k_\vdash \rangle_{AA'} = \frac{1}{\sqrt{d}} \sum_{l = 0}^{d-1} e^{\frac{2 \pi i}{d} k l} |j,l \rangle_{AA'}.\label{eq:MUB}
\end{equation}
\textcolor{black}{We will denote the test measurement with the notation $\{\hat{\mathcal{T}}(jk)\}$, where the effect is defined as $\hat{\mathcal{T}}(jk) \rho = |j,k_\vdash \rangle\langle j,k_\vdash |$, occuring with probability $\langle j,k_\vdash |\rho|j,k_\vdash \rangle$.}

Thus, $c = \max_{j,k,x,y} | \langle j k_{\vdash} | \mathcal{B} (xy) \rangle |^2$. To estimate $c$, we notice that
\begin{eqnarray}
   \nonumber  \langle j k_{\vdash} | \mathcal{B} (xy) \rangle && = \frac{1}{d} \sum_{l,m = 0}^{d-1} e^{-\frac{2 \pi i}{d}(km - ly)} \langle j,m | l,l \oplus x \rangle \\
 \nonumber  && = \frac{1}{d} \sum_{l,m = 0}^{d-1} e^{-\frac{2 \pi i}{d}(km - ly)} \delta_{j,l} \delta_{m, l \oplus x} \\
 && = \frac{1}{d} e^{-\frac{2 \pi i}{d}(k(j \oplus x) - jy)}, \label{eq:c_value}
\end{eqnarray}
which directly implies $c = 1/d^2 ~ \forall~ (jk), (xy)$. Thus, the key rate in this protocol is given by $r \geq \log_2 d^2$, provided that the shared state is maximally entangled.

\textbf{Remark $\mathbf{1}$.} We have considered the maximally entangled state and all the measurements in the computational basis. If, instead, we are working with some other basis, i.e., if the maximally entangled state is given by $\ket{\psi} = \frac{1}{\sqrt{d}} \sum_{i = 0}^{d - 1} \ket{e_i f_i}$, the measurements, too, need to be rotated in order to obtain the proper key rate. Note that there always exist local unitaries $U$ and $V$, such that $U \ket{e_i} = \ket{i}$ and $V \ket{f_j} = \ket{j}$, where $\{|i\rangle\} ~ \text{or} ~\{|j\rangle\}$ represents the computational basis of the corresponding Hilbert spaces. In such a case, the auxiliary maximally entangled state $\ket{\phi^+}_{B' C}$ (which is concatenated with the original resource for the encoding process in the purified protocol) and the Bell-basis measurements $\{|\mathcal{B}(xy)_{BB'}\rangle\}$ require to be rotated through $V \otimes V$, whereas the decoding Bell-measurement should be unitarily rotated by $U \otimes V$. A similar procedure needs to be adopted for the encoding and decoding test measurements as well. This would ensure that the key rate for every maximally entangled state, considered in any arbitrary basis, amounts to $2 \log_2 d$. Therefore, in our proof, we shall only consider pure states in the computational basis, along with all subsequent measurements. Note that, since it is \textit{a priori} allowed in the key distribution protocol to know the exact resource state, the parties may as well rotate it to the computational basis before the start of the protocol, which is technically easier than rotating the aforementioned measurements. This is necessary to obtain the proper key rate since we keep the measurements in our protocol fixed to the ones elucidated in the previous section.

\section{Secret key rate in the ideal SDC protocol}
\label{sec:pure-SDC}

In this section, we derive the key rates for non-maximally entangled resource states employed in the SDC protocol. We consider Bell-diagonal mixed states of different ranks in $d \otimes d$. Furthermore, only the shared states are considered to be mixed, whereas the quantum channels through which the initial and the encoded qudits are sent are assumed to be noiseless. Note that, if one considers pure states throughout the protocol, it means that there are no quantum correlations created with the eavesdropper. However, there are attacks like intercept-resend strategies and symmetric individual attacks~\cite{Gisin_RMP_2002}, which Eve can avail in order to obtain information about the transmitted states. Such attacks can be detected during the test runs, and the honest parties can choose to abort the protocol. On the other hand, in the case of mixtures of Bell states, the eavesdropper's presence cannot be ruled out. With this in mind, we obtain a sufficient condition for Bell mixtures to furnish a positive key rate which, surprisingly, connects the Holevo capacity (insecure rate of classical information transmission) and the secure key rate. Before presenting our results, let us briefly describe the dense coding capacity.

\textit{Dense coding capacity}. For a single sender ($B$) and a single receiver ($A$), the dense coding capacity is defined as $C \equiv C(\rho_{AB}) = \log_2 d_B  + S(A)_\rho - S(AB)_\rho$~\cite{Ziman_PRA_2003, Bruss_PRL_2004, Bruss_IJQI_2006}. $d_B$ is the dimension of the sender's subsystem with $C_{\text{cl}} = \log_2 d_B$ being the classical threshold. A state provides quantum advantage in DC (\textcolor{black}{non-classicality in terms of DC})  when $S(A)_\rho - S(AB)_\rho > 0$. Since $S(AB)_\rho = 0$ for pure states, pure entangled states are always beneficial for DC. 

\subsection{Connecting key rates with DC capacity for bipartite Bell mixtures}
\label{subsubsec:noiseless-mixed}

For a given dimension $d$, any bipartite mixed state can have a rank, $R \leq d^2$. Let us consider bipartite mixed states of rank $R$ as the convex mixture of $R$ Bell states, $\ket{\mathcal{B}(xy)}$, as defined in Eq.~\eqref{eq:high_d_Bell}, i.e., $\rho_{AB}^R = \sum_{x = 0}^{x_d-1} \sum_{y = 0}^{d_x-1} p^{xy} \ket{\mathcal{B}(xy)} \bra{\mathcal{B}(xy)}$, with $\sum_{x, y} p^{xy} = 1$. 
\textcolor{black}{Throughout this paper, we frequently refer to Bell mixtures, so it is important to note a subtlety regarding the state $\rho_{AB}^R = \sum_{x,y} p^{xy} \ket{\mathcal{B}(xy)} \bra{\mathcal{B}(xy)}$. Being diagonal in the $d$-dimensional Bell basis, the eigenvalues of the state $\rho_{AB}^R$ are given by $\{p^{xy}\}$, and its rank is thus determined by the number of non-vanishing $p^{xy}$. Since $x,y \in \{0, d-1\}$, for every $x$, we can have $d$ different Bell states, characterized by a specific value of $y$, which differ from each other through a local phase. Let us suppose that in the index $x$, there are total $x_d$ different values of $x$, and for each $x$, there are $d_x$ different $y$ values for which $p^{xy}$ is non-zero. Hence, the rank of  $\rho_{AB}^R$ is given by $R = \left(\sum_{x = 0}^{x_d-1} \left( \sum_{y = 0}^{d_x-1} 1 \right) \right)$.}
\textcolor{black}{It is worth mentioning that the individual elements in the convex sum of $\rho_{AB}^R$, i.e.,  
$\ket{\mathcal{B}(xy)}$, (given in Eq.~\eqref{eq:high_d_Bell}) reduce to the well known Bell states for $d = 2$.
}
Note that this choice of mixed states is a suitable candidate for the protocol, as preparing pure Bell states is an ideal scenario and the presence of either Eve or noise in the first channel $\mathcal{E}_{A' \to B}$, naturally leads to Bell mixtures. In what follows, we shall see that choosing such states has a profound impact on the key rate that can be obtained via the SDC protocol. \textcolor{black}{But first, let us note that for the mixed state $\rho_{AB}^R$, the dense coding capacity is given by $C = \log_2 d + S(\rho_A^R) - S(\rho_{AB}^R) = 2 \log_2 d + \sum_{x,y} p^{xy} \log_2 p^{xy} =  C_{\text{cl}} + \log_2 d  + \sum_{x, y} p^{xy} \log_2 p^{xy}$,  where the local density matrix, $\rho_A^R = \mathbb{I}/d$, is a maximally mixed state having maximum von Neumann entropy $\log_2 d$. Therefore, in order to have quantum advantage, or some non-classical capacity,} we must demand that $\sum_{x, y} p^{xy} \log_2 p^{xy} > -\log_2 d$. We now present our main result for mixed states employed in the SDC protocol.


\textbf{Theorem $\mathbf{1}$.} \textit{A Bell mixture of rank $R~(\leq d^2)$ in $\mathbb{C}^d\otimes\mathbb{C}^d$ provides a positive key rate if its dense coding capacity is non-classical.}

\textit{Proof}. We consider $\rho_{AB}^R = \sum_{x = 0}^{x_d-1} \sum_{y = 0}^{d_x-1} p^{xy} \ket{\mathcal{B}(xy)} \bra{\mathcal{B}(xy)}$ as our resource in the SDC protocol, where $\ket{\mathcal{B}(xy)}$ are given by Eq.~\eqref{eq:high_d_Bell}, \textcolor{black}{with rank, $(\sum_{x = 0}^{x_d-1} (\sum_{y = 0}^{d_x-1} 1)) = R$}. Here, the second summation represents the $d_x$ Bell states $\ket{\mathcal{B}(xy)}$ which have the same value of $x$ but varying values of $y$, and the first summation indicates the $x_d$ number of sets $\{\ket{\mathcal{B}(xy)}\}$ which differ in their parameter $x$. 
The conditional entropy after the key generation run is given as $S(B|A)_\kappa = S(AB)_\kappa - S(A)_\kappa = (\log_2 d^2 - \sum_{x,y} p^{xy} \log_2 p^{xy}) - \log_2 d^2 = - \sum_{x, y} p^{xy} \log_2 p^{xy}$, whereas corresponding to the test run $S(B|A)_\tau = S(AB)_\tau - S(A)_\tau = \left(\log_2 d^2 -\sum_{x = 0}^{x_d-1} (\sum_{y = 0}^{d_x-1} p^{xy}) \log_2 (\sum_{y = 0}^{d_x-1} p^{xy})\right) - \log_2 d^2 = -\sum_{x = 0}^{x_d-1} (\sum_{y = 0}^{d_x-1} p^{xy}) \log_2 (\sum_{y = 0}^{d_x-1} p^{xy})$ (see Appendix~\ref{app:1}). Then, the secure key obtainable from this state is bounded below by Eq.~\eqref{eq:key_rate_bound} as
\begin{eqnarray}
   \nonumber  r && \geq 2\log_2d + \sum_{x = 0}^{x_d-1} \sum_{y = 0}^{d_x-1} p^{xy} \log_2 p^{xy} \\\nonumber && + \sum_{x = 0}^{x_d-1} (\sum_{y = 0}^{d_x-1} p^{xy}) \log_2 (\sum_{y = 0}^{d_x-1} p^{xy}) ~~~~ \\
    && \geq \log_2 d + \sum_{x = 0}^{x_d-1} (\sum_{y = 0}^{d_x-1} p^{xy}) \log_2 (\sum_{y = 0}^{d_x-1} p^{xy}), \label{eq:theorem2_proof1}
\end{eqnarray}
where in the second inequality we have used the fact that $\sum_{x = 0}^{x_d-1} \sum_{y = 0}^{d_x-1} p^{xy} \log_2 p^{xy} \geq -\log_2 d$, which ensures quantum advantage in the dense coding capacity. We now define $\tilde{p}^x = \sum_{y = 0}^{d_x-1} p^{xy}$ which satisfies $\sum_{x = 0}^{x_d-1} \tilde{p}^x = 1$. Evidently, $\{\tilde{p}^x\}$ also forms a valid probability distribution, since each $\tilde{p}^x$ is a sum of positive numbers $p^{xy}$, which are therefore positive, satisfying the normalization condition. Therefore, we can rewrite Eq.~\eqref{eq:theorem2_proof1} as
\begin{eqnarray}
\label{eq:bell_mixture_keyrate_bound}
    r \geq \log_2 d + \sum_{x = 0}^{x_d-1} \tilde{p}^x \log_2 \tilde{p}^x.
\end{eqnarray}
Given that the $x \in \{0, d - 1\}$, we may estimate $\sum_{x = 0}^{x_d-1} \tilde{p}^x \log_2 \tilde{p}^x \geq -\log_2 d$. Therefore, $r \geq 0$, whenever the mixed state can provide quantum advantage in the dense coding routine. Hence the proof. $\hfill \blacksquare$

We will now derive the maximum and minimum values of the right-hand side (RHS) of Eq. (\ref{eq:bell_mixture_keyrate_bound}) for Bell mixtures in $d \otimes d$ having rank $R$, provided that they are useful in the dense coding protocol. Let us proceed by considering three different cases. In the following, for brevity, we shall use RHS to denote the right-hand side of an equation.

\underline{\textbf{Case $\mathbf{1}$.} $R \leq d$}: In case the mixed state has a rank less than the dimension of one of its subsystems, it trivially follows that $S(B|A)_\kappa = - \sum_{x = 0}^{x_d-1} \sum_{y = 0}^{d_x-1} p^{xy} \log_2 p^{xy} \leq \log_2d$ since $(\sum_{x = 0}^{x_d-1} (\sum_{y = 0}^{d_x-1} 1)) \leq d$. In order to obtain the maximum RHS, we choose all the $R$ Bell states to have the same value of $x$, i.e., $x_d = 1$ and $d_x = R-1$, which implies, $S(B|A)_\tau = - (\sum_{y = 0}^{R-1} p^y) \log_2 (\sum_{y = 0}^{R-1} p^y) = 0$, since all the probabilities involved in the convex mixture must sum up to unity. Therefore, $\text{RHS}_{\max}^{R \leq d} = \log_2 d$. On the other hand, if we choose each of the $R$ states to have different values of $x$, implying $x_d = R-1$ and $d_x = 1$, we obtain $S(B|A)_\tau = - \sum_{x = 0}^{R-1} p^x \log_2 p^x \leq \log_2 R$. Furthermore, the RHS is minimized when $p^x = 1/R~\forall~x$, and $\text{RHS}_{\min}^{R \leq d} \geq \log_2 \frac{d}{R}$. 

\underline{\textbf{Case $\mathbf{2}$.} $R > d$}: In this scenario, we have to impose $S(B|A)_\kappa \leq \log_2 d$ from the condition of quantum advantage in the dense coding protocol, since $- \sum_{x = 0}^{x_d-1} \sum_{y = 0}^{d_x-1} p^{xy} \log_2 p^{xy} \geq \log_2d$ in general. Let $R = nd + d'$ with $1 \leq n < d$, and $d' < d$. The maximum RHS is obtained when, for each of the $n$ cases, we choose all the $d$ Bell states to have the same value of $x = x_n$ and the remaining $d'$ such states to also be characterized by the same $x \neq x_n$. Then, we can write $S(B|A)_\tau = - \sum_{x = 0}^{n-1} \tilde{p}^x \log_2 \tilde{p}^x - \tilde{p}^{d'} \log_2 \tilde{p}^{d'} \leq \log_2 n - \tilde{p}^{d'} \log_2 \tilde{p}^{d'}$, where $\tilde{p}^{d'} = \sum_{y = 0}^{d'-1} p^{(x = d')y}$ corresponds to the probabilities with which the last $d'$ Bell states are mixed. Therefore, $\text{RHS}_{\max}^{R > d} \geq \log_2 \frac{d}{n}  - \tilde{p}^{d'} \log_2 \tilde{p}^{d'}$. The RHS obtains its minimum value when the probabilities in the mixture are taken in such a way that the minimum number of Bell states possess the same $x$ and $S(B|A)_\tau$ is maximized. One way is to set $p^{xy} = 1/R$ for all the states, whence $\text{RHS}_{\min}^{R >d} \geq \log_2 d - \frac{n d}{R} \log_2 \frac{d}{R} - \frac{d'}{R} \log_2 \frac{d'}{R}$.

\underline{\textbf{Case $\mathbf{3}$.} $R = d^2$} : When the mixed state is of full rank, $p^{xy} = \frac{1}{d^2}$ for all states gives the  minimum RHS. In such a scenario, $S(B|A)_\tau = -\sum_{x = 0}^{d-1} \tilde{p}^x \log_2 \tilde{p}^x = \log_2 d$ since $\tilde{p}^x = \sum_{y = 0}^{d-1} p^{xy} = 1/d$. Thus, $\text{RHS}_{\min}^{R = d^2} \geq 0$. To maximize the RHS for the full rank state, we must choose the mixing probabilities in such a way that $S(B|A)_\tau$ is minimized. Note that in this case too, $S(B|A)_\kappa \leq \log_2 d$ from the dense coding advantage. 

\begin{table}[]
	\caption{Average regularized key rate.} 
	\begin{tabular}{|r|rrrr|}
\hline
\multicolumn{1}{|c|}{$d$} & \multicolumn{4}{c|}{$\langle \tilde{r} \rangle$}                                                                                    \\ \hline
\multicolumn{1}{|c|}{}    & \multicolumn{3}{c|}{Bell mixture}                                                              & \multicolumn{1}{c|}{Random states} \\ \hline
\multicolumn{1}{|l|}{}    & \multicolumn{1}{c|}{$R = 2$}  & \multicolumn{1}{c|}{$R = 3$}  & \multicolumn{1}{c|}{$R = 4$}   & \multicolumn{1}{c|}{$R = 2$}       \\ \hline
2                         & \multicolumn{1}{r|}{0.180636} & \multicolumn{1}{r|}{0.105261} & \multicolumn{1}{r|}{0.0989495} & 0.288065                           \\ \hline
3                         & \multicolumn{1}{r|}{0.483039} & \multicolumn{1}{r|}{0.12709}  & \multicolumn{1}{r|}{0.10578}   & 0.362568                           \\ \hline
4                         & \multicolumn{1}{r|}{0.590318} & \multicolumn{1}{r|}{0.308235} & \multicolumn{1}{r|}{0.12253}   & 0.459374                           \\ \hline
5                         & \multicolumn{1}{r|}{0.647119} & \multicolumn{1}{r|}{0.404146} & \multicolumn{1}{r|}{0.226962}  & 0.516064                           \\ \hline
\end{tabular}
	\label{tab:key_avg}
\end{table}

\subsection*{The regularized key rate and dimensional advantage}

It may seem trivial that as $d$ increases, so does the lower bound on the key rate $r$. To better emphasize the constructive effect of using higher dimensional systems as resources in the SDC protocol, let us regularize the key rate with the corresponding maximum value, i.e., we define $\tilde{r} = \frac{r}{2\log_2d}$. Note that $\tilde{r} = 1$, when the shared state is $\ket{\phi^+}$ in any dimension. However, the dimensional advantage on the noiseless SDC protocol becomes apparent, when we consider the shared states to be mixed states. We illustrate our claim through Table~\ref{tab:key_avg}, where we enumerate $\langle \tilde{r} \rangle = \frac{\tilde{r}}{N}$ -  the average of the regularized key rate over $N$ simulated mixed states. Our data indicates that for Bell mixtures of rank, $R = 2$ to $4$, higher dimensional resource states provide a higher $\tilde{r}$ and, therefore, a better key rate. To emphasize the advantage further, we also simulate random states with $R = 2$, in which case too, the advantage persists upon increasing $d$.

\textbf{Note $\mathbf{1}$.} We  simulate $N = 10^4$ Bell mixtures of  rank, $R = 2, 3, 4$ as well as the same number of rank-$2$ random states for our numerical analysis. In case of the Bell mixtures, $\rho_{AB}^R = \sum_{x,y} p^{xy} \ket{\mathcal{B}(xy)} \bra{\mathcal{B}(xy)}$, we randomly generate values of $x, y \in \{0, 1, \cdots, d - 1\}$ from a uniform distribution to create the $R$ orthogonal Bell states according to Eq.~\eqref{eq:high_d_Bell}, \textcolor{black}{and corresponding probabilities, $p^{xy}$s, are chosen from uniform distribution, satisfying $\sum_{x,y} p^{xy}=1$.} In the case of random states, we generate the convex mixture $\tilde{\rho}_{AB}^2 = p^1 \ket{\psi^1}\bra{\psi^1} + p^2 \ket{\psi^2}\bra{\psi^2}$, with $\ket{\psi^1} = \sum_{i = 0}^{d - 1} \alpha_i \ket{ii}$ and $\ket{\psi^2} = \sum_{i = 0}^{d - 1} \alpha'_i \ket{i(i \oplus 1)}$, where the values of $\alpha_i, \alpha'_i$ are randomly chosen from a Gaussian distribution of vanishing mean and unit standard deviation. Choosing the values of $p^i \in (0,1)$ randomly from a uniform distribution, satisfying $\sum_i p^i = 1$, we calculate the average lower bound on the key rate according to the purified protocol in Sec.~\ref{sec:noiseless_high}.

\section{Dimensional advantage in noisy SDC  protocol}
\label{sec:noisy_qkd}

We here derive the key rates when the shared state, which we consider to be the maximally entangled bipartite state, $\ket{\phi^+} = \frac{1}{\sqrt{d}}\sum_{i = 0}^{d - 1} \ket{ii}$ in arbitrary dimensions, is subjected to some paradigmatic noise models. Note that the noise affects the protocol twice - once when Alice shares the state with Bob, and again when Bob sends the encoded particle to Alice. Therefore, the key rate is not expected to remain positive for high noise strengths. We examine the range of the noise parameter, $p$, within which the shared state can still provide an advantage in the key distribution, and analyze how the key rate varies with the dimension $d$. 
We now proceed to derive the key rate in the presence of three exemplary noise models - the dit-phase-flip, the depolarising, and the amplitude-damping channels.

\subsection{Noisy SDC key rates}
\label{subsec:weyl-noise}



We consider the higher dimensional realizations of three well-known noise models. For $d \geq 3$, the Kraus operators characterizing such noise models are again given in terms of the Weyl operators as in Eq.~\eqref{eq:unitary}~\cite{Fonesca_PRA_2019}. Specifically, the depolarising, the dit-phase-flip, and the amplitude-damping channels are mathematically represented, respectively, by 
\begin{eqnarray}
 && \Lambda_{\text{dp}} : \hat{K}^{00} = \sqrt{1 - \frac{d^2 - 1}{d^2}p} \hat{U}^{00}; ~~ \hat{K}^{ij} = \frac{p}{\sqrt{d - 1}} \hat{U}^{ij} ~~~~~~~~ \\
   && \nonumber ~~~~~~~~~~~~~~~~~~\text{with}~ 0 \leq i,j \leq d - 1 ~ \text{for} ~ (i,j) \neq (0,0), \\
    && \Lambda_{\text{d-ph}} : \hat{K}^{00} = \sqrt{1 - p} \hat{U}^{00}; ~~ \hat{K}^{ij} = \frac{\sqrt{p}}{d - 1} \hat{U}^{ij} \\
   && \nonumber ~~~~~~~~~~~~~~~~~~~~~~~~~~~~~\text{with}~ 1 \leq i,j \leq d - 1, \\
   &&  \Lambda_{\text{ad}}:  \hat{K}^0 = \ket{0}\bra{0} + \sqrt{1 - p} \sum_{i = 1}^{d - 1} \ket{i}\bra{i}; \hat{K}^i = \sqrt{p} \ket{0} \bra{i} ~~~~~~ \\
    &&  \nonumber ~~~~~~~~~~~~~~~~~~~~~~~~~~~~~ \text{with}~ 1 \leq i \leq d - 1,
\end{eqnarray}
where $0 \leq p \leq 1$ is the noise parameter.

\begin{figure*}
    \includegraphics[width=\linewidth]{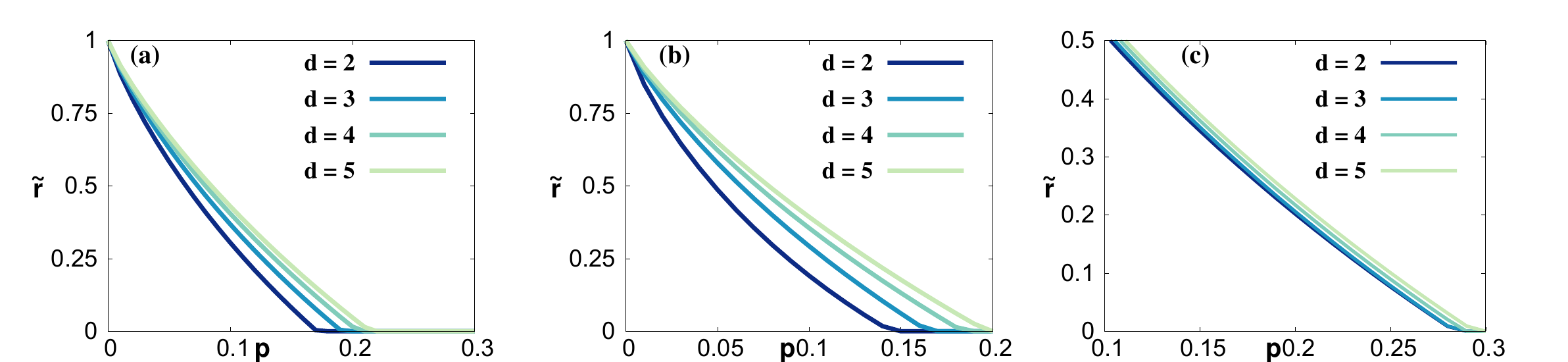}
     \caption{\textbf{\textcolor{black}{Regularised key rate for $\ket{\phi^+}$ against the noise strength.} } The regularized key rate $\tilde{r}$ (ordinate) against the noise strength $p$ (abscissa) when the pure maximally entangled state $\ket{\phi^+}$ is employed in the noisy SDC protocol affected by (a) depolarising noise, (b) dit-phase-flip noise, and (c) amplitude-damping noise. Dimensions $d = 2, 3, 4, 5$ are represented through dark to light lines respectively. The y-axis is in bits, whereas the x-axis is dimensionless.}
    \label{fig:key}
\end{figure*}

 \textcolor{black}{The expressions for the key rates achieved by any of the Bell states affected by the depolarising and dit-phase-flip noises are given by}
\begin{widetext}
    \begin{eqnarray}
       \nonumber  r_{\text{dp}} \geq && 3 \log_2 \left(d^2\right)+\frac{(d-1)^2 p^2 \log_2 \left(\frac{p^2}{d^4}\right)+2 (d-1) p (d-(d-1) p) \log_2 \left(\frac{p (d-(d-1) p)}{d^4}\right)+(d-(d-1) p)^2 \log_2 \left(\frac{(d-(d-1) p)^2}{d^4}\right)}{d^2} \\
       && -\frac{\left(1-d^2\right) (2-p) p \log_2 \left(\frac{(2-p) p}{d^4}\right)-\left(d^2-\left(d^2-1\right) (2-p) p\right) \log_2 \left(\frac{d^2-\left(d^2-1\right) (2-p) p}{d^4}\right)}{d^2 }, \label{eq:depo_rate} \text{and} \\
       \nonumber r_{\text{d-ph}} \geq && 3 \log_2 \left(d^2\right) + p \left(p \log_2 \left(\frac{p^2}{(d-1)^2 d^2}\right)-2 (p-1) \log_2 \left(-\frac{(p-1) p}{(d-1) d^2}\right)\right)+(p-1)^2 \log_2 \left(\frac{(p-1)^2}{d^2}\right) \\
       && \nonumber -\frac{-2 (d-2) p^2 \log_2 \left(\frac{(d-2) p^2}{(d-1)^3 d^2}\right)+p \left(\left(d^2-2\right) p-2 (d-1)^2\right) \log_2 \left(\frac{p \left(2 (d-1)^2-\left(d^2-2\right) p\right)}{(d-1)^4 d^2}\right)}{(d-1)^2} \\
       && - \frac{\left(p \left(2 (d-1)^2-\left((d-1)^2+1\right) p\right)-(d-1)^2\right) \log_2 \left(\frac{p \left(\left((d-1)^2+1\right) p-2 (d-1)^2\right)+(d-1)^2}{(d-1)^2 d^2}\right)}{(d-1)^2} \label{eq:dph-rate}. 
    \end{eqnarray}
\end{widetext}
\textcolor{black}{Eqs.~\eqref{eq:depo_rate} and~\eqref{eq:dph-rate} provide a lower bound on the key rate furnished by the state $\frac{1}{\sqrt{d}}\sum_{i = 0}^{d - 1} \ket{i,i}$, when affected by depolarising and dit-phase-flip noises, in terms of the noise parameter $p$ and the corresponding dimension $d$. Note that, in our calculations, we explicitly calculate the bounds for $d = 2, 3, 4$ which allow us to recursively reach the final expressions for arbitrary $d$.}
We skip the expressions for the key rate under the influence of amplitude-damping noise since the compact mathematical form is too cumbersome.

\begin{figure}
    \includegraphics[width=\linewidth]{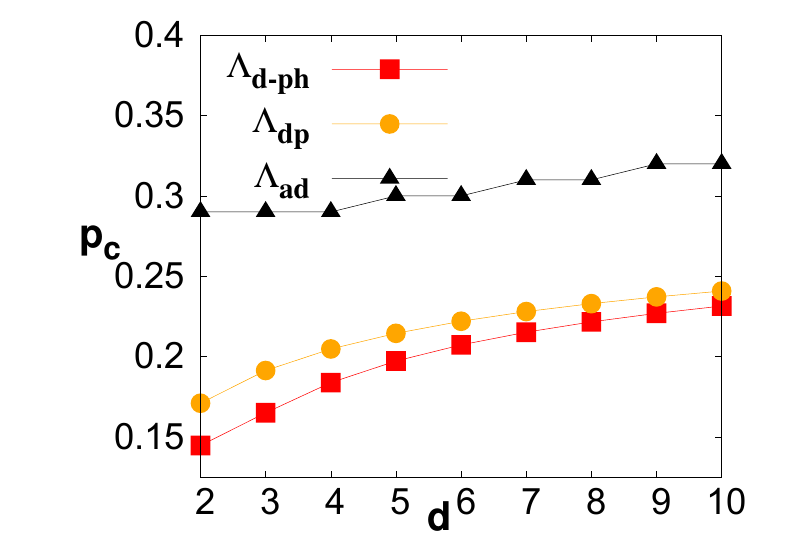}
     \caption{\textbf{Critical noise strength vs dimension. } The variation of the critical noise strength $p_c$ (ordinate) is illustrated against the dimension $d$ (abscissa) of the noisy $\ket{\phi^+}$ state. The squares, circles, and triangles represent the dit-phase-flip, the depolarising, and the amplitude-damping noise models, respectively. Both axes are dimensionless. }
    \label{fig:pc}
\end{figure}

\begin{figure*}
    \includegraphics[width=\linewidth]{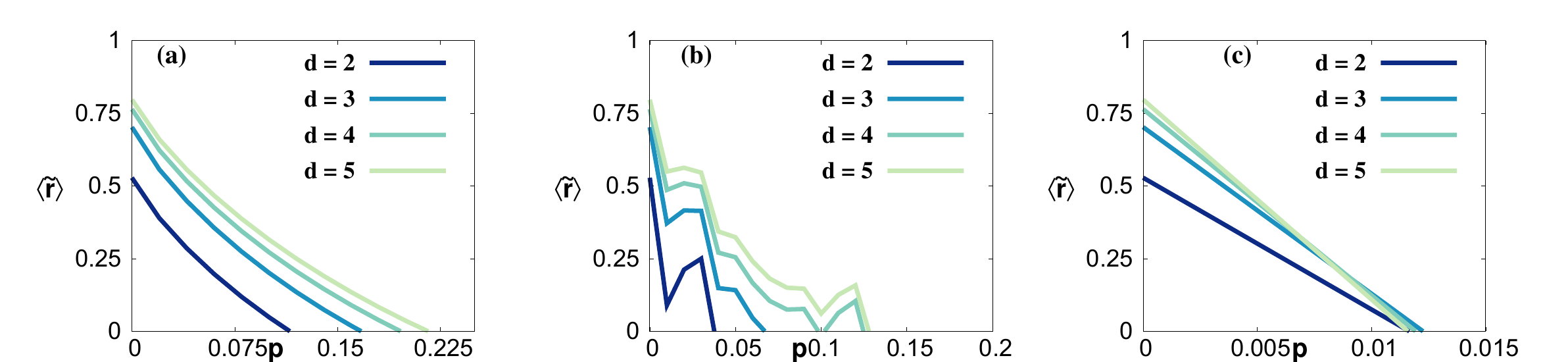}
     \caption{\textbf{\textcolor{black}{Average regularised key rate of rank-$2$ Bell mixture against the noise strength.} } The average regularized key rate $\langle \tilde{r} \rangle$ (ordinate) is plotted against the noise strength $p$ (abscissa) when $10^4$ rank-$2$ Bell mixtures, \textcolor{black}{$\rho_{AB}^R = \sum_{x,y} p^{xy} \ket{\mathcal{B}(xy)} \bra{\mathcal{B}(xy)}$ with $R=2$, in  dimension $d\otimes d$ (with $d=2,3,4, \text{and }5$), are generated and then used in the noisy SDC protocol under (a) depolarising noise, (b) dit-phase-flip noise, and (c) amplitude-damping noise.} All other specifications are the same as in Fig.~\ref{fig:key}. The horizontal axis is dimensionless while the vertical one is in bits.   }
    \label{fig:key-mix}
\end{figure*}

\subsection{ Dimensional benefits with noisy channels }
\label{subsubsec:d-adv}

In the presence of noise, we report two advantages of this protocol -- $(1)$ the dimensional advantage persists even in the noisy scenarios of the SDC scheme, and $(2)$ the key rate remains non-vanishing for moderate values of the noise strength $p$, which demonstrates the robustness of the key distribution protocol. In particular, let us highlight the variation of the regularized key rate $\tilde{r}$ with respect to the noise parameter $p$, thereby establishing the two-fold advantage in increasing the dimension.
\begin{enumerate}
    \item \textbf{Dimensional gain survives with noise. } When $p = 0$, the secret key rate achieves its maximum value, i.e., $\tilde{r} = 1$ when the shared state is a maximally entangled state. As expected, $\tilde{r}$ decreases with the increase of $p$ irrespective of the dimension (see Fig.~\ref{fig:key}). However, for a given $p$, $\tilde{r}^{(d_1)} > \tilde{r}^{(d_2)}$, when $d_1 > d_2$. This illustrates that the increase in the key rate with dimension persists even in the presence of noise. Interestingly, we observe that the dimensional gain depends on the nature of the noise. In particular, the key rate is higher when the amplitude-damping noise is in effect, as compared to the Weyl-type noise models. This weaker destructive effect of the amplitude-damping noise comes at the cost of smaller dimensional advantage, i.e., the increase in the key rate with $d$ is smaller than that obtained in the case of the dit-phase-flip and depolarising noise models, i.e., at a given $p$, we have $\tilde{r}_{\text{dp}(\text{d-ph})}^{(d_1)} - \tilde{r}_{\text{dp}(\text{d-ph})}^{(d_2)} > \tilde{r}^{(d_1)}_{\text{ad}} - \tilde{r}^{(d_2)}_{\text{ad}}$ with $d_1 > d_2$.

    \item \textbf{Resilience against noise. } Another crucial impact of increasing $d$ stems from the fact that the critical noise strength, $p_c$, at which the key rate vanishes, also increases with dimension. This implies that the protocol becomes more robust against environmental effects as the resource dimension is increased. Figure~\ref{fig:pc} depicts how $p_c$, at which $\tilde{r} = 0$, increases steadily with $d$, although for the amplitude-damping noise, the noise robustness is much less pronounced compared to the Weyl-type noise models. Surprisingly, the depolarising noise, which is known to affect communication protocols and quantum correlations much more severely~\cite{Mishra_PLA_2015, Dutta_PLA_2016, Fonesca_PRA_2019, Gupta_PRA_2022, Muhuri_arXiv_2022, Dutta_PS_2023} exhibits a higher value of $p_c$ than the one in the case of the dit-phase-flip noise model, although, $p_c^{(d_1)} - p_c^{(d_2)}$ (with $d_1 > d_2$) is greater for $\Lambda_{\text{d-ph}}$ than for $\Lambda_{\text{dp}}$, i.e., the increase in $p_c$ with dimension is higher for the dit-phase-flip channel. This possibly indicates that the key distribution protocol is influenced by noise in a different way than other information-theoretic protocols.

    \textcolor{black}{Note that the noise parameter, $p$, is related to the time duration, $t$, of the interaction of the transmitted qudits with the environment, such that a higher value of $p$ indicates a higher interaction time~\cite{nielsenchuang_CUP_2000, Lidar_arXiv_2019}. Therefore, an increased noise resilience, i.e., a higher range of $p$ for which a positive key rate is obtained, translates into a longer distance for the successful implementation of the SDC protocol. As a result, the use of qudits imparts the benefit of the key distribution over longer distances compared to the qubit-based protocol.}
\end{enumerate}

\textcolor{black}{\textbf{Remark.} We can also interpret the enhanced noise resilience with respect to the stability of the channel, i.e., the extent to which it transmits the qudits perfectly. Due to interaction with the environment, the independent use of separate channels leads to greater decoherence in comparison with the one-way QKD protocols~\cite{Gisin_RMP_2002}. Since we consider noise both before and after encoding and establish how qudits can offer better noise resilience, our results suggest that the SDC protocol with higher dimensional systems can also aid in stabilizing the involved channels for a longer time.}

\subsection*{Effect of noise on the SDC protocol with Bell mixtures}
\label{subsec:noisy_Bell-mix}

\textcolor{black}{Let us now study the noisy SDC protocol when the mixture of Bell states,  $\rho_{AB}^R = \sum_{x,y} p^{xy} \ket{\mathcal{B}(xy)} \bra{\mathcal{B}(xy)}$, is shared between Alice and Bob. We choose rank-$2$ Bell mixtures (i.e., $R=2$) in dimension $d\otimes d$ with $d = 2, 3, 4, 5$ in this case.} For every dimension, we randomly simulate $10^4$ such states, as explained in Note $1$ earlier, and calculate the mean regularized key rate $\langle \tilde{r} \rangle$ at each value of the noise strength, $p$. The dimensional advantage is ubiquitous in terms of the higher key rate furnished by the Bell mixture as $d$ increases (as illustrated in Fig.~\ref{fig:key-mix}). The resistance to noise is apparent only in the case of Weyl noises, i.e., for the depolarising and the dit-phase-flip noises, where the value of $p_c$ increases with dimension. For the amplitude-damping noise, however, the critical noise strength (when the key rate vanishes) actually decreases very slightly with $d$, thereby signifying a destructive effect of dimensional increment. This occurs possibly due to the influence of the non-unital nature of the amplitude-damping noise present in the channels.
On the other hand, the effect of the amplitude-damping noise on the magnitude of the key rate is much lower as compared to the Weyl-noise scenario, and contrary to the noiseless case, the dimensional advantage on $\tilde{r}$ is more pronounced in the noisy scenario.

\textcolor{black}{\textbf{Remark.} It is evident from Fig.~\ref{fig:key-mix} that the average value of $\langle \tilde{r} \rangle$ demonstrates non-monotonic behavior under the dit-phase-flip noise strength, contrasting with the depolarizing and amplitude-damping noise scenarios. This can be due to the fact that the states are not generated randomly based on the Haar measure~\cite{bengtsson_zyczkowski_2006}. Given that a particular noise model affects the Bloch (hyper)sphere in a distinct way, the uneven distribution of states within the (hyper)sphere could significantly contribute to this non-monotonic behavior.}

\section{The set of non-resourceful states for the SDC protocol}
\label{sec:set}

We now focus our attention on the set of states that are useless for key generation through the SDC protocol, i.e., the states for which $r < 0$. Since it has been established that non-dense codeable states form a convex and compact set~\cite{Vempati_PRA_2021}, it will be intriguing to study whether the corresponding set of states employed in the SDC protocol also does so. 

\textbf{Theorem $\mathbf{2}$.} \textit{The states useless for the key generation in the SDC protocol form a convex and compact set.}

\textit{Proof}. We shall prove the statement in two parts.

\textbf{Convexity.} Let $\rho_1$ and $\rho_2$ be two states acting on $d \otimes d$ which provide a non-positive key rate, i.e.,
\begin{eqnarray}
    && \nonumber 2 \log_2 d - S(B|A)_{\rho_i, \kappa} - S(B|A)_{\rho_i, \tau} <0 \\
   && \implies S(B|A)_{\rho_i, \kappa} + S(B|A)_{\rho_i, \tau} > 2\log_2 d,
    \label{eq:th4-eq1}
\end{eqnarray}
where $i = 1,2$ while $\kappa$ and $\tau$ refer to the states in Eqs.~\eqref{eq:cc-state} and~\eqref{eq:tau_state} respectively. Let us consider $\rho = p_1 \rho_1 + p_2 \rho_2$ with $p_1 + p_2 = 1$. Since the conditional entropy is a concave function of its arguments, one obtains
\begin{eqnarray}
  \nonumber && S(B|A)_{\rho, \kappa(\tau)} \geq p_1 S(B|A)_{\rho_1, \kappa(\tau)} + p_2 S(B|A)_{\rho_2, \kappa(\tau)} \\
 \nonumber  && \implies S(B|A)_{\rho, \kappa} + S(B|A)_{\rho,\tau} \geq \sum_{i = 1}^2 p_i \Big(S(B|A)_{\rho_i, \kappa} + \\
 \nonumber && ~~~~~~~~~~~~~~~~~~~~~~~~~~~~~~~~~~~~~~~~~~~~~~~~~~~~~~~~~~~~~~~~S(B|A)_{\rho_i, \tau}\Big) \\
 && \implies S(B|A)_{\rho, \kappa} + S(B|A)_{\rho,\tau} \geq 2 \log_2 d.
    \label{eq:th4-eq2}
\end{eqnarray}
Therefore, any convex mixture of useless states in the SDC protocol also yields a useless state.

\textbf{Compactness.} Note that the set of states useless in the SDC protocol is bounded since every valid state in a finite-dimensional Hilbert space admits a bounded spectrum (the eigenvalues always lie between $0$ and $1$). Secondly, $S(B|A)_{\kappa (\tau)}$ vanishes for any maximally entangled state, whereas it attains a value of $2 \log_2 d$ for the maximally mixed state in $d \otimes d$. Therefore, in $d$-dimension, $S(B|A)_{\kappa} + S(B|A)_{\tau} = [0, 4 \log_2 d]$. Thus, the set of non-resourceful states for the key generation using the SDC protocol is bounded, and since the conditional entropy is a continuous function~\cite{Alicki_JPA_2004, Winter_CMP_2016}, it is also closed, thereby implying that it is compact. Hence the proof.~$\hfill \blacksquare$

\section{Conclusion}
\label{sec:conclu}

The dense coding protocol (DC) has been well studied with respect to the transfer of classical information over a quantum channel and has also been realized in laboratories. Hence, the preparation of secure keys based on DC is both theoretically interesting as well as experimental-friendly. Moreover, it is deterministic and provides a higher key rate than the well-known two-way BB84 protocol, while still maintaining its advantage against losses encountered in fiber-optical platforms. 

In this work, the semi-device independent security proof of the secure dense coding (SDC) protocol was generalized to higher dimensions when the shared states are chosen to be different classes of mixed states. We derived the lower bound on the achievable key rate through the purification approach, which assumes that the malicious party can distribute the resource among the actual participants and, therefore, considers the most general forms of eavesdropper attacks. Our analysis clearly specified the encoding, decoding, and test-measurement schemes required to implement the protocol using two-qudit states. We established a connection between the Holevo capacity and the key rate, thereby showing that the secure and insecure rates of classical information transmission are connected for a specific class of states. Formally, we found that convex mixtures of orthogonal Bell states can furnish a positive key rate when their Holevo capacity exhibits quantum advantage in the noiseless scenario. We also demonstrated the dimensional advantage of producing a higher key rate when mixed two-qudit states are employed.

In recent times, one of the main challenges that any quantum information processing task suffers from is decoherence. Since the systems cannot be perfectly isolated, they always interact with the environment, which is responsible for reducing the quantum features inherent in the system, thereby affecting the performance of the protocol. To address this issue, we considered three exemplary models - the depolarising, the dit-phase-flip, and the amplitude-damping channels and their respective effects on the secret key rate. We identified two distinct benefits of using higher dimensional systems in the presence of noise, one concerning a higher key rate for a given noise strength, and the other being the increased robustness against noise. We manifested this analytically when the shared state is a two-qudit maximally entangled state, while a similar behavior can also be observed via numerical simulation of Bell mixtures. Finally, we proved that the states useless in the secure dense coding protocol form a convex and compact set. This indicates that one can construct witness operators to identify states viable for the protocol. 

\textcolor{black}{It is interesting to note that entanglement-free versions of entanglement-based QKD protocols have been proposed, which offer similar secure key rates without the expense of creating entanglement. Notably, the protocol proposed in Ref.~\cite{Lucamarcini_PRL_2005}, referred to as the LM05 scheme, serves as the entanglement-free version of the ping-pong protocol~\cite{Bostrom_PRL_2002}. The original version of the ping-pong protocol, however, was formulated with the purpose of secure transmission at the expense of dense coding and involved the transmission of only one qubit leading to the generation of one bit of secure key, which, in the absence of entanglement, was replicated by the LM05 protocol. On the other hand, the entanglement-based SDC protocol considered in this work enables the creation of two dits of secure key through the transmission of only one qudit due to the presence of the additional resource of a pre-shared entangled state. Therefore, if an entanglement-less version of the same is proposed, then transferring the same amount of information would require double the number of qudits to be transferred, against the norm of entanglement-based and entanglement-free QKD protocols having the same underlying structure. The transmission of only one qudit at any given run would reduce the key string by half in the corresponding protocol without entanglement. As such, the formulation of an entanglement-free version of the SDC protocol is rather non-trivial and we leave its security analysis as a direction for future research.}

In conclusion, let us state that the advantage of using higher dimensional systems in different quantum key distribution protocols has already been established. Our work again emphasizes the constructive effect of higher dimensional Hilbert spaces on the implementation of deterministic two-way cryptographic schemes.

\section*{Acknowledgement} 
AP, RG, and ASD acknowledge the support from the Interdisciplinary Cyber-Physical Systems (ICPS) program of the Department of Science and Technology (DST), India, Grant No.: DST/ICPS/QuST/Theme- 1/2019/23.  We acknowledge the use of \href{https://github.com/titaschanda/QIClib}{QIClib} -- a modern C++ library for general-purpose quantum information processing and quantum computing (\url{https://titaschanda.github.io/QIClib}).  This research was supported in part by the 'INFOSYS scholarship for senior students'.

\appendix
\textcolor{black}{
\section{Derivation of conditional entropy terms in Theorem $1$.}
\label{app:1}
In Theorem $1$, we consider a rank $R$ Bell mixture with just a change of the dummy variables as $\rho_{AB}^R = \sum_{\alpha = 0}^{x_d-1} \sum_{\beta = 0}^{d_x-1} p(\alpha,\beta) \ket{\mathcal{B}(\alpha \beta)} \bra{\mathcal{B}(\alpha \beta)}$, with $\ket{\mathcal{B}(\alpha\beta)}$ defined in Eq.~\eqref{eq:high_d_Bell}. In order to derive $S(B|A)_\kappa$, we first concatenate $\rho_{AB}^R$ with $\frac{1}{\sqrt{d}} \sum_{i = 0}^{d - 1} \ket{i,i}_{B'A'}$ and then perform Bell measurements $\ket{\mathcal{B}(rs)}$ on the parties $B, B'$. Alice further performs another Bell measurement $\ket{\mathcal{B}(ij)}$.
Let us consider the post-measurement classical-classical \footnote{The state of the form $\rho_{AB}^c = \sum_{x,y = 1}^d p(x,y)\ket{x}\bra{x}_A \otimes \ket{y}\bra{y}_B$, with $\sum_{x,y = 1}^d p(x,y) = 1$, is called a classical-classical state as it does not contain any kind of quantum correlation. All kind of quantum correlation measures turn out to be zero for this state, for example $\mathcal{D}(\rho_{AB}^c) = 0$, where $\mathcal{D}$ is a bipartite quantum correlation measure, known as quantum discord~\cite{discord_review}. } state in part of Alice and Bob as
\begin{eqnarray}
  && \nonumber \kappa_{AA'BB'} \\
\nonumber  &&= \sum_{i,j,r,s} \hat{\mathcal{B}}_{AA'}(ij) \otimes \hat{\mathcal{B}}_{BB'}(rs) (\rho_{AB}^R \otimes \ket{\phi^+}\bra{\phi^+}_{A'B'}) \\
    && = \sum_{i,j,r,s} g(ijrs) |i j\rangle_{AA'} \langle i j| \otimes |rs\rangle_{BB'}\langle rs|,
    \label{eq:ccq-state_Appendix}
\end{eqnarray}
where $g(ijrs)$ is given by 
\begin{widetext}
\begin{eqnarray}
   g(ijrs) &=& {}\bra{\mathcal{B}(rs)}_{BB'} {}\bra{\mathcal{B}(ij)}_{AA'}\left(\rho_{AB}^R \otimes \ket{\phi^+}\bra{\phi^+}_{A'B'}\right)\ket{\mathcal{B}(ij)}_{AA'}\ket{\mathcal{B}(rs)}_{BB'} \\
   &=&\sum_{\alpha = 0}^{x_d-1} \sum_{\beta = 0}^{d_x-1} p(\alpha,\beta)  {}\bra{\mathcal{B}(rs)}_{BB'} {}\bra{\mathcal{B}(ij)}_{AA'}\left(\ket{\mathcal{B}(\alpha \beta)} \bra{\mathcal{B}(\alpha \beta)}_{AB} \otimes \ket{\phi^+}\bra{\phi^+}_{A'B'}\right)\ket{\mathcal{B}(ij)}_{AA'}\ket{\mathcal{B}(rs)}_{BB'} \\
   &=& \frac{1}{d^2} \sum_{\alpha = 0}^{x_d-1} \sum_{\beta = 0}^{d_x-1} p(\alpha,\beta) \delta_{i, \alpha \oplus r} \delta_{j, \beta \oplus s},
\end{eqnarray}
\end{widetext} 
and $g(ij) = \sum_{r,s} g(ijrs) = 1/d^2$.
The Shannon entropy corresponding to $\{g(ijrs)\}$ gives us $S(AB)_\kappa$ while $S(A)_\kappa = H(\{ij\})$ with \(H(\{p_i\}) = - \sum_i p_i \log_2 p_i\). This helps us to find $S(B|A)_\kappa$ as described in the proof of Theorem $1$.\\
A similar analysis, but with the Bell measurements on $BB'$ and $AA'$ replaced by $\mathcal{T}$ as defined in Eq.~\eqref{eq:MUB}, allows us to derive $S(B|A)_\tau$. 
\begin{eqnarray}
    &&\tau_{AA'BB'} \nonumber \\
    &&= \sum_{i,j,x,y} \hat{\mathcal{T}}_{AA'}(ij) \otimes \hat{\mathcal{T}}_{BB'}(rs) (\rho_{AB}^R \otimes \ket{\phi^+}\bra{\phi^+}_{A'B'}) , \nonumber \\ 
    && = \sum_{ijrs} h(ijrs) |i j\rangle_{AA'} \langle i j| \otimes |rs\rangle_{BB'}\langle rs|, \label{eq:tau_state_Appendix}
\end{eqnarray}
where $h(ijrs)$ is given by 
\begin{widetext}
    \begin{eqnarray}
        h(ijrs)   &=& \bra{r s_{\vdash}}_{BB'} {}\bra{i j_\vdash}_{AA'}\left(\rho_{AB}^R \otimes \ket{\phi^+}\bra{\phi^+}_{A'B'}\right)\ket{ij_\vdash}_{AA'}\ket{rs_\vdash}_{BB'} \\
   &=&\sum_{\alpha = 0}^{x_d-1} \sum_{\beta = 0}^{d_x-1} p(\alpha,\beta)  \bra{r s_{\vdash}}_{BB'} {}\bra{i j_\vdash}_{AA'}\left(\ket{\mathcal{B}(\alpha \beta)} \bra{\mathcal{B}(\alpha \beta)}_{AB} \otimes \ket{\phi^+}\bra{\phi^+}_{A'B'}\right)\ket{ij_\vdash}_{AA'}\ket{rs_\vdash}_{BB'} \\
   &=& \frac{1}{d^2} \sum_{\alpha = 0}^{x_d-1} \sum_{\beta = 0}^{d_x-1} p(\alpha,\beta) \delta_{i, \alpha \oplus r} \delta_{js}.
    \end{eqnarray}
\end{widetext}
Once again, we have the marginals $h(ij) = \sum_{rs}h(ijrs) = \frac{1}{d^2}$, and we can calculate the entropies in terms of the probabilities $h(ijrs)$.
After deriving the conditional entropy terms, we arrive at the expressions presented in Theorem $1$.}

\bibliography{Ref}
\bibliographystyle{apsrev4-1}

\end{document}